\documentclass{nature}
\usepackage{eqnarray}
\usepackage[english]{babel} 
\usepackage[]{layout}
\usepackage{amsmath,amsfonts,amssymb}
\usepackage{graphicx}
\usepackage{float}
\usepackage{bm}
\usepackage{color}
\usepackage{braket}
\usepackage{array,multirow,makecell}
\makeatletter
\let\saved@includegraphics\includegraphics
\AtBeginDocument{\let\includegraphics\saved@includegraphics}
\renewenvironment*{figure}{\@float{figure}}{\end@float}
\makeatother

\title{Angular-dependent Klein tunneling in photonic graphene}

\author{Zhaoyang Zhang$^{1\ast}$, Yuan Feng$^{1}$, Feng Li$^{1\ast}$, Sergei Koniakhin$^{2,3}$,  Changbiao Li$^{1}$, Fu Liu$^{1}$, Yanpeng Zhang$^{1}$, Min Xiao$^{4,5}$, Guillaume Malpuech$^{2\ast}$, Dmitry Solnyshkov$^{2,6\ast}$}

\begin{document}
\maketitle

\begin{affiliations}
\item Key Laboratory for Physical Electronics and Devices of the Ministry of Education \& Shaanxi Key Lab of Information Photonic Technique, School of Electronic and Information Engineering, Faculty of Electronic and Information Engineering, Xi'an Jiaotong University, Xi'an 710049, China
\item Institut Pascal, PHOTON-N2, Universit\'e Clermont Auvergne, CNRS, SIGMA Clermont, F-63000 Clermont-Ferrand, France.
\item Center for Theoretical Physics of Complex Systems, Institute for Basic Science (IBS), Daejeon 34126, Republic of Korea
\item Department of Physics, University of Arkansas, Fayetteville, Arkansas, 72701, USA
\item National Laboratory of Solid State Microstructures and School of Physics, Nanjing University, Nanjing 210093, China
\item Institut Universitaire de France (IUF), 75231 Paris, France
\end{affiliations}

\begin{abstract}
The Klein paradox consists in the perfect tunneling of relativistic particles through high potential barriers\cite{Klein1929,Greiner1985}. As a curious feature of particle physics, it is responsible for the exceptional conductive properties of graphene \cite{Katsnelson2006}. It was recently studied in the context of atomic condensates \cite{Salger2011} and topological photonics and phononics \cite{Dreisow2012,Jiang2020}. While in theory the perfect tunneling holds only for normal incidence \cite{Katsnelson2006}, so far the angular dependence of the Klein tunneling and its strong variation with the barrier height were not measured experimentally. In this work, we capitalize on the versatility of atomic vapor cells with paraxial beam propagation and index patterning by electromagnetically-induced transparency \cite{Zhang2019}. We report the first experimental observation of perfect  Klein transmission in a 2D photonic system (photonic graphene) at normal incidence and measure the angular dependence. Counter-intuitively, but in agreement with the Dirac equation, we observe that the decay of the Klein transmission versus angle is suppressed by increasing the barrier height, a key result for the conductivity of graphene and its analogues.  
\end{abstract}

\maketitle


The Klein paradox was initially discovered in the beginning of the XX century \cite{Klein1929}, immediately after the formulation of the Dirac equation. The ensuing impossibility for a relativistic electron to be confined inside the nucleus was a very important conclusion, which ultimately led to the discovery of the neutron \cite{Chadwick1932}. The description of the Klein tunneling involves the quantum field theory and the concept of particles and antiparticles \cite{Greiner1985,Dombey1999,Krekora2004}. However, the perfect Klein tunneling  with relativistic electrons has never been observed experimentally, because it requires extremely high and strongly-localized energy barriers \cite{Greiner1985}.

The studies of the Klein tunneling have shown a new surge of interest in the XXI century with the advent of analogue systems, such as graphene \cite{Semenoff1984,Katsnelson2006,DasSarma2011}, cold atoms \cite{Salger2011}, photonics \cite{Dreisow2012}, and even acoustic systems \cite{Jiang2020}, where the Dirac Hamiltonian can be simulated on-demand. The first crucial result was that the exceptional conductivity of graphene is due precisely to this phenomenon \cite{Katsnelson2006,Huard2007,Stander2009}. Further studies were focused on the \emph{energy dependence} of the tunneling with respect to the barrier height in different systems~\cite{Jiang2020}.

On the other hand, the \emph{angular dependence} of the Klein tunneling has never been studied experimentally. The tunneling is expected to be perfect only for normal incidence, and theoretically, it should  quickly drop with the angle, depending on the particle energy and on the barrier height \cite{Katsnelson2006,Allain2011}. This dependence is crucial both from the fundamental point of view, and for the applied properties of graphene \cite{Katsnelson2006} and its optical analogues, as the incidence is rarely perfectly normal in realistic situations. 

Photonic graphene has recently emerged as a promising analogue system with various implementations \cite{Rechtsman2013b,Plotnik2014,lu2014topological}. It is based on the propagation of a light probe beam in a 2D honeycomb lattice of refractive index. The propagation of the probe beam is well described by the paraxial approximation of Maxwell's equations, where the propagation along an axis is mapped to an effective time (see Methods). This configuration allows to emulate various Hamiltonians, including the one of graphene \cite{Ozawa2019}. The advantage of optical systems is the possibility of direct observation of the wavefunctions, instead of the integral quantities used in solid state physics, such as conductivity. For example, in topological photonics it has allowed the observation of the photonic quantum Hall \cite{Haldane2008,Wang2009,Rechtsman2013b,Hafezi2013} and quantum spin or valley Hall effects \cite{Wu2015,Ma2015,ma2016all,Xu2016,Barik2016,gao2018topologically,Zhang2018}, with the associated chiral edge states, promising for applications. Recently, reconfigurable photonic graphene in atomic vapor cells has allowed to study the dynamics of singularities (quantum vortices) appearing in the Dirac equation \cite{Zhang2019}.

In this work, we use the photonic graphene implemented in an atomic vapor cell to study the angular and barrier height dependence of the Klein tunneling. Our measurements are in excellent agreement with the analytical theory based on the Dirac equation. We observe a perfect tunneling for normal incidence and a strong decrease of the transmission versus the incidence angle. This angular dependence is much weaker if the barrier height is increased. This never observed counter-intuitive behavior is probably responsible for the high conductivity of electronic graphene, where scattering defects typically correspond to large potential barriers. 

\begin{figure}[tbp]
\centering
\includegraphics[width=0.9\linewidth]{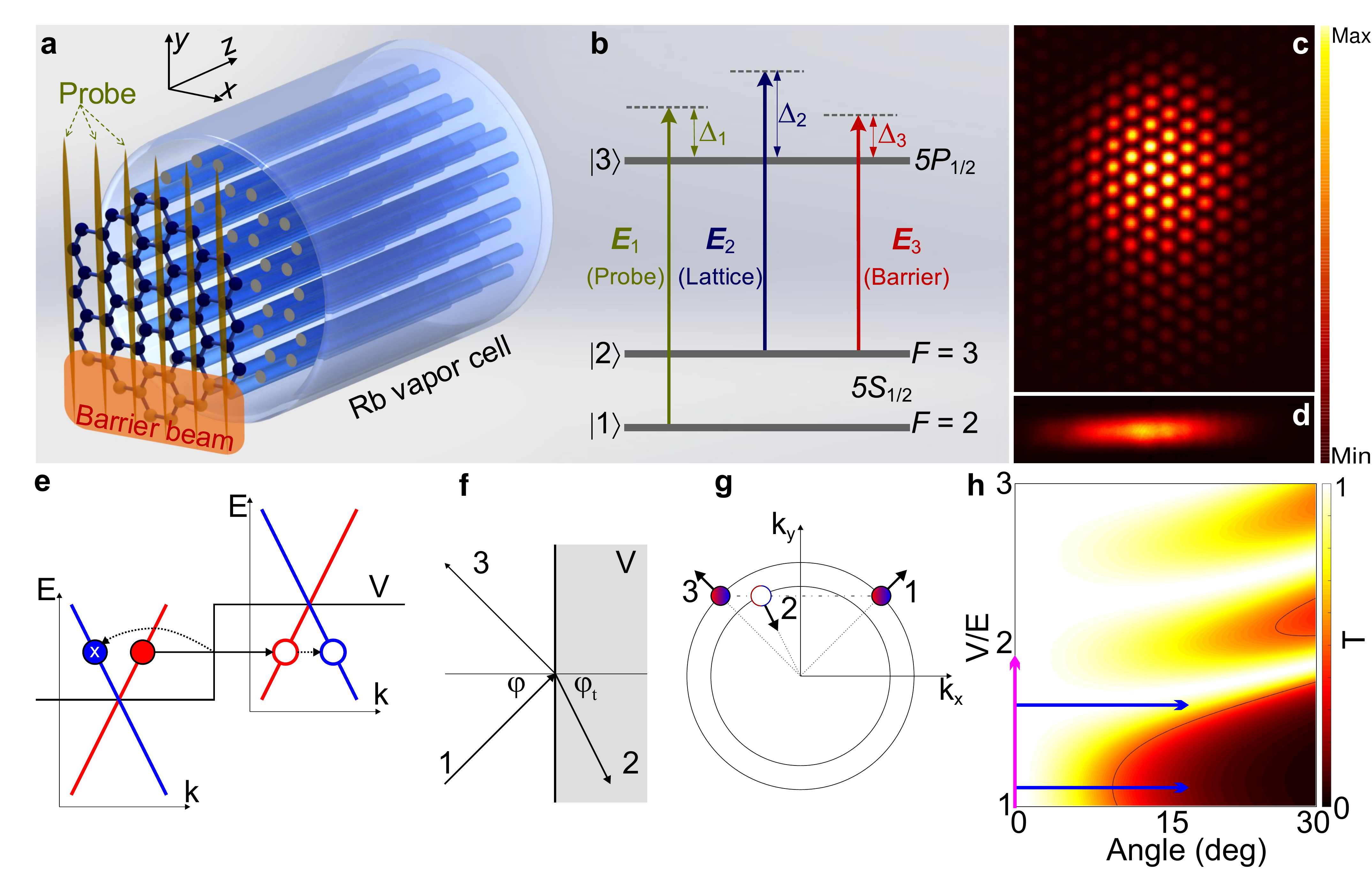}
\caption{\small \textbf{Photonic graphene and Klein tunneling.} \small \textbf{a-c} \textbf{a} A honeycomb lattice is formed in the Rb vapor cell by the interference of three coupling beams \bm{$E_2$}, \bm{$E'_2$} and \bm{$E''_2$}. The probe beam, formed by the interference of \bm{$E_1$} and \bm{$E'_1$}, is  close either to the Dirac or $\Gamma$ point. The potential barrier is formed by another stripe-shaped coupling beam \bm{$E_3$}. All beams interact with each other via the EIT effect. \textbf{b} The Rb atomic energy levels and the configuration of the EIT effect.  \textbf{c} Experimental image of the interference patterns formed by the three coupling beams (\bm{$E_2$}, \bm{$E'_2$} and \bm{$E''_2$}): the dark sites form a honeycomb lattice. \textbf{d} Experimental image of the barrier beam. \textbf{e-g} Scheme of the Klein tunneling: filled circles -- particle states, empty circles -- hole states. \textbf{e} Perfect Klein tunneling in 1D. The color of lines of the dispersion ($E(k)$) and circles (states) correspond to the conserved particle-hole pseudospin. Crosses mark the forbidden final states. $V$ is the potential. \textbf{f} Klein tunneling in 2D (real space), arrows indicate the group velocity. \textbf{g} States in the 2D reciprocal space. $k_y$ is conserved. Arrows mark the group velocity. All transitions are allowed, because the states are not orthogonal. \textbf{h} Transmission through a finite barrier calculated with Eq.~\eqref{KAngle} as a function of the incidence angle and the barrier height. Magenta and blue arrows: experimental scans (Figs.~2 and 3). \label{fig1}}
\end{figure}


To study the Klein tunneling and its peculiar angular dependence, we used a  highly-re\-con\-fi\-gu\-rable implementation of photonic graphene based on electromagnetically-induced transparency (EIT) in atomic vapors~\cite{Zhang2019,Zhang2020,Zhang2020b}. The scheme of the experimental setup is shown in Fig.~\ref{fig1}a. Three vertically polarized coupling beams \bm{$E_2$}, \bm{$E'_2$} and \bm{$E''_2$} from the same external-cavity diode laser propagate along z with a small angle $\sim$ $0.5^{\circ}$ between each other, forming a hexagonal interference pattern in the x-y plane (Fig.~\ref{fig1}c). The photonic graphene is formed by the dark lattice sites which exhibit higher refractive index, as determined by the chosen two-photon detuning (Fig.~\ref{fig1}b). Formed by the interference of the coupling beams, the honeycomb pattern stays approximately uniform along $z$ throughout the entire length of the atomic cell. A narrow, line-shaped beam  \bm{$E_3$} (from a second laser) parallel to the $z$-axis serves as a static potential barrier (see Fig.~\ref{fig1}d, Methods). To excite the Dirac points, the probe beam is split into two beams \bm{$E_1$} and \bm{$E'_1$}, forming an interference pattern that covers only the A or B sub-lattice, controlling the pseudospin at the Dirac points \cite{Zhang2019}. The probe feels the combined modulation of susceptibility imposed by the EIT effects of both the lattice and the barrier beams. To observe the Klein tunneling, the probe beams are moved slightly away from the Dirac point by fine-adjusting the beam angle. The probe beams have a combined momentum along $-y$ and interact with the honeycomb lattice and the barrier, while all of them are propagating along $z$ through the atomic vapor cell. In this configuration, the 2D dynamics of Klein tunneling occurs within the $x-y$ plane, and the $z$ axis is regarded as the effective axis of time (see Methods). Similarly, we can send a probe beam close to the $\Gamma$ point for comparison.

The Klein tunneling is best known in the case of a massless Dirac Hamiltonian for ultra-relativistic particles. The model system for the study of the angular dependence of the tunneling should be at least two-dimensional. The corresponding Hamiltonian reads:
\begin{equation}
    \hat{H}=c\bm{\sigma}\cdot\hat{\bm{p}}
    \label{Dirham}
\end{equation}
where $\hat{\bm{p}}$ is a two-dimensional momentum operator and $\bm{\sigma}$ is a vector of Pauli matrices.  This effective Hamiltonian is implemented at the corners of the Brillouin zone in honeycomb lattices \cite{Semenoff1984}. The corresponding eigenstates are characterized by a linear dispersion of eigenenergies $E=\pm\hbar c |k|$ and a particle-antiparticle spinor for eigen-states $(\pm e^{-i\varphi},1)^T/\sqrt{2}$, where $\varphi$ is the polar angle of the wave vector $\bm{k}$. The perfect Klein tunneling under normal incidence is due to the conservation of this pseudospin and to the particle-hole conversion, making possible the propagation at negative energies. This is illustrated in Fig.~\ref{fig1}e, showing in the normal incidence case with the dispersion branches in two regions characterized by potentials 0 and $V$. The color of the branches indicates the configuration of the spinor  $(-1,1)^T/\sqrt{2}$ (red) and $(1,1)^T/\sqrt{2}$ (blue). At the barrier, the incident particle is converted into a hole with a negative energy, which continues to propagate in the same direction because it has the same group velocity. Backscattering (from ``red" to ``blue" states) is impossible because of the pseudospin conservation.

In 2D, for an arbitrary angle of incidence, the spinors of the branches involved in the scattering are not orthogonal anymore, and the reflection becomes allowed. It is illustrated in Fig.~\ref{fig1}f. The angle of transmission $\varphi_t$ is described by an analogue of the Snell-Descartes law,  $E\sin\varphi=-(E-V)\sin\varphi_t$. It can be either higher or lower  than the angle of incidence, depending on the barrier height $V$ with respect to the particle energy $E$. The scheme of the process in the reciprocal space is shown in Fig.~\ref{fig1}f,g: the wavevector along the interface $k_y$ is conserved. For $E>V$, the isoenergetic circle for the holes is smaller. In this case, the transmission at high angles becomes completely impossible and the reflection is total. For a barrier of a  finite length $d$, the transmission can be found analytically as\cite{Allain2011}:
\begin{equation}
    T=\frac{\cos^2\varphi\cos^2\varphi_t}{\cos^2\varphi\cos^2\varphi_t\cos^2(k_x'd)+\sin^2(k_x'd)(1+\sin\varphi_t\sin\varphi)^2}
    \label{KAngle}
\end{equation}
where $k_x'd=-2\pi l\sqrt{1-2\epsilon+\epsilon^2\cos^2\varphi}$, $l=Vd/(2\pi\hbar c)$, $\epsilon=E/V$.  This expression exhibits a strong angular and barrier height dependence. In particular, the angular dependence is maximal when $V/E\to 1$, where only a narrow range of incident angles lead to transmission, with total external reflection at higher angles. This transmission is shown in Fig.~\ref{fig1}h as a function of the angle of incidence and the barrier height. In the following, we present the experimental scan of this dependence along the arrows (magenta and blue) shown in this panel.

\begin{figure}[tbp]
\centering
\includegraphics[width=0.5\linewidth]{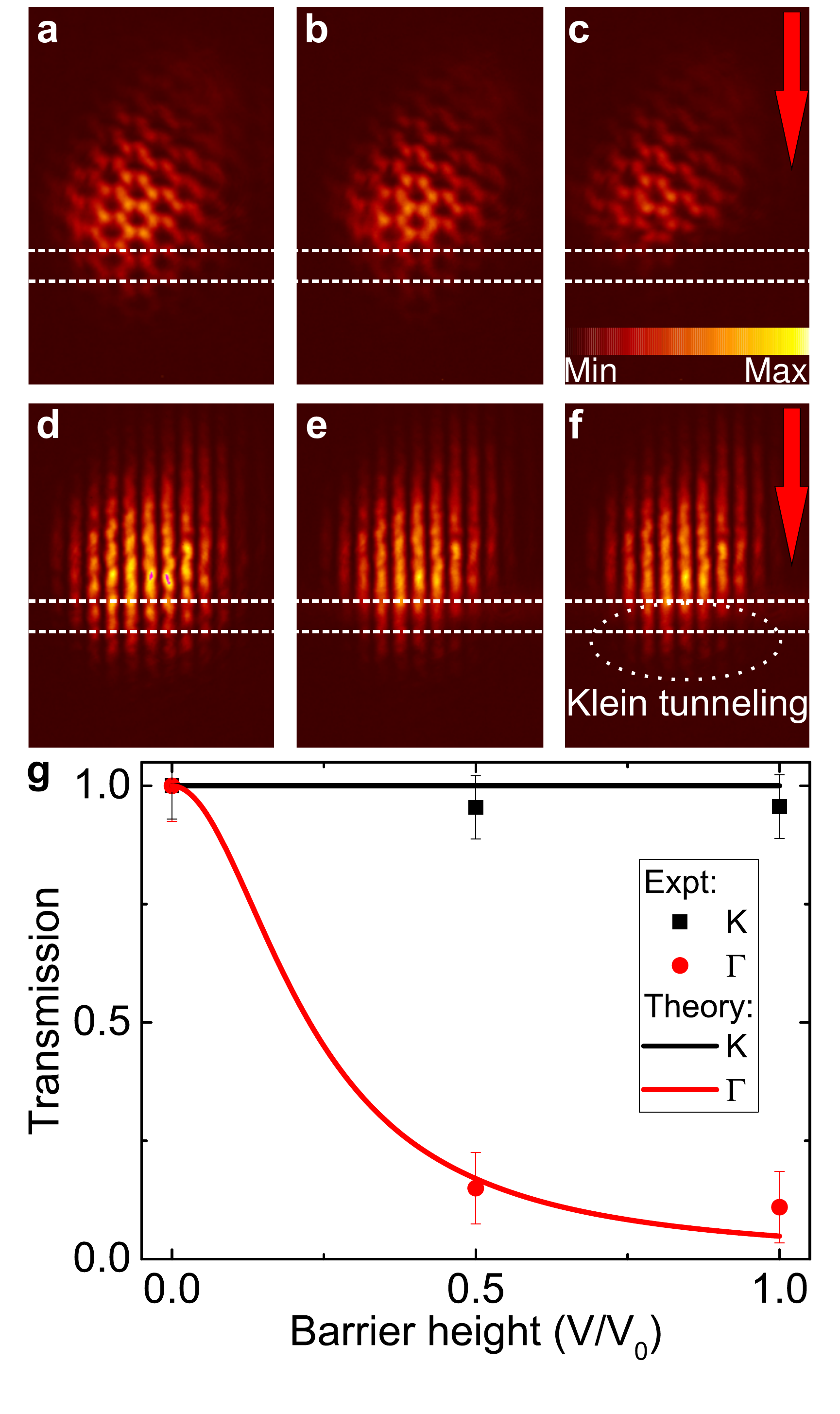}
\caption{\textbf{Klein tunneling at normal incidence.} The barrier is delimited by white dashed lines. The probe motion is indicated by the red arrows. \textbf{a-c} Ordinary quantum-mechanical tunneling at the $\Gamma$ point. \textbf{d-f} Klein tunneling. The high transmission region is highlighted by a dashed ellipse. 
\textbf{g} Transmission for both cases (dots -- experiment, lines -- theory) for $K$ and $\Gamma$ points. Error bars mark the experimental uncertainty. \label{fig2}}
\end{figure}

First, we show the experimental results confirming the existence of the Klein tunneling at normal incidence (magenta arrow in Fig.~\ref{fig1}h), comparing it with the case of massive non-relativistic particles as a reference. Figure~\ref{fig2} shows the experimental images of ordinary quantum-mechanical tunneling (panels a-c), which is achieved at the $\Gamma$ point in photonic graphene. The barrier height is increased from left to right by experimentally changing the frequency detuning of the barrier beam, and the transmitted intensity clearly decreases. For the case of Klein tunneling, achieved with a wave packet close to the K point described by the Dirac Hamiltonian~\eqref{Dirham}, the increase of the barrier height does not lead to obvious decrease of the tunneling (panels d-f). Even for the highest barrier ($V=V_0$, see Methods), the intensity can still be observed inside and beyond the barrier region (marked ``Klein tunneling" in the figure). This is summarized in Fig.~\ref{fig2}g, showing the relative transmission with zero-barrier baseline (see Methods) as a function of the barrier height in both cases, as well as the theoretical predictions. In the case of Klein tunneling (probe close to the K point), a full transmission $T=1$, independent of the barrier height $V$ is expected theoretically (black solid line). This is indeed confirmed by the experiment (black squares). This full transmission is in striking contrast with the behavior of the wave packets close to the $\Gamma$ point, whose transmission exhibits a rapid decrease with the barrier height in experiments (red circles). This case is well described by the usual quantum-mechanical tunneling formula for massive particles (red solid line).
\begin{figure}[tbp]
\centering
\includegraphics[width=0.9\linewidth]{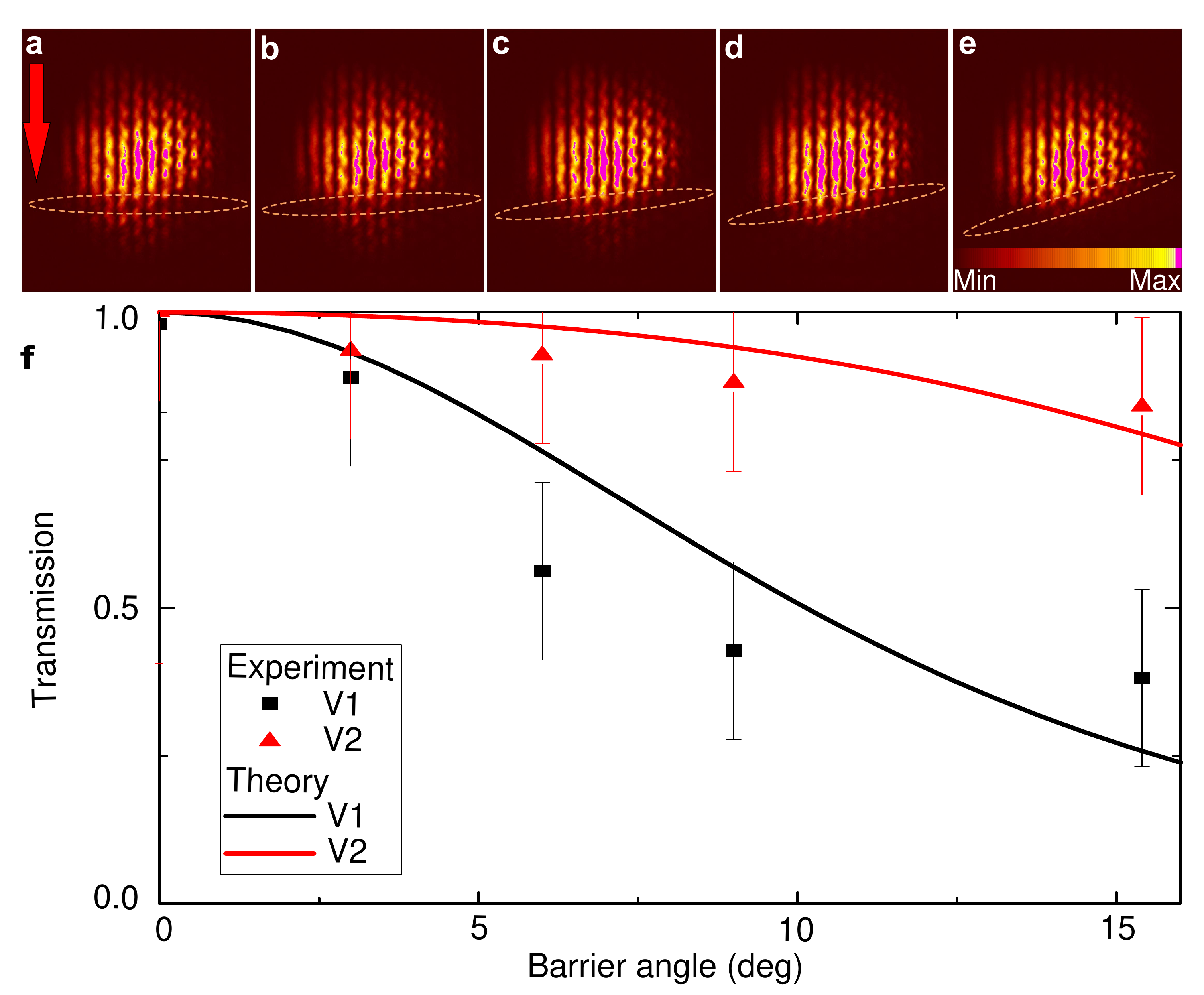}
\caption{\textbf{Angular and barrier height-dependent Klein tunneling.} The barrier position is marked with a dashed ellipse. The probe motion is marked by the red arrow. \textbf{a-e} Experimental images of the probe intensity after the cell with barrier rotation angles $0$, $3$, $6$, $9$, and $15^\circ$, respectively.  \textbf{f} Angular dependence of the transmission for two barrier heights ($V1$ -- black, $V2$ -- red, $V1<V2$): dots -- experiment, lines -- theory. The error bars mark the experimental uncertainty. $V1$ and $V2$ corresponds to a detuning $\Delta_3$=-90MHz and -110MHz, respectively.   \label{fig3}}
\end{figure}

Finally, we demonstrate the strong angular dependence of the Klein tunneling (Fig.~\ref{fig3}) for weak barrier height and the suppression of this dependence when the barrier becomes higher (two blue arrows in Fig.~\ref{fig1}h). Panels~a-e show experimental images of the wavepacket with varying angle of the barrier potential (from 0 to $\sim$15 degrees) with respect to the probe wave vector. The experimentally measured transmission is plotted in Fig.~\ref{fig3}f as black squares and red triangles. The two sets of data correspond to different barrier heights $V_1\approx 1.1E$, $V_2 \approx 1.7E$ (see Methods).  Theoretical results, calculated using Eq.~\eqref{KAngle}, are shown as solid lines of the corresponding color. The good agreement of the experiment and the theory in Fig.~\ref{fig3} confirms the validity of the description of graphene-like structures in terms of the Dirac Hamiltonian.  Overall, a higher tunneling is observed at all angles for a \emph{higher} barrier $V_2>V_1$. A smaller barrier exhibits a faster decrease of the tunneling with the angle of incidence. This counter-intuitive behavior can be understood qualitatively from the modified Snell-Descartes law illustrated by Fig.~\ref{fig1}g: reducing the barrier height reduces the transmitted wave vector and extends the total external reflection region for $V/2<E<V$. Thus, the Klein tunneling is paradoxical not only in its perfectness, but also in its rapid decay with incidence angle for small barriers. One can expect, for example, that smaller defects could affect the transport properties stronger than more pronounced ones.


To conclude, we have used an implementation of photonic graphene based on EIT to measure one of the key features of the Klein tunneling -- its strong and counter-intuitive angular dependence. Our work makes an important step towards the potential applications of the relativistic particle-hole physics of graphene-like structures.

\begin{methods}

\textit{Paraxial beam propagation.} 
Our work is based on the equivalence between the paraxial approximation of the Helmholtz equation for the electric field of a stationary electromagnetic wave in a dielectric media and the 2D time-dependent Schr\"odinger equation,  where the propagation axis for the electric field maps to the time axis of the quantum-mechanical problem. The Helmholtz equation reads:
\begin{equation}
    \nabla^2\bm{E}^2+k_0n^2\bm{E}=0
\end{equation}
where $k_0$ is the wave vector and $n$ is the refractive index. The paraxial approximation means considering the envelope of the electric field and neglecting $\partial^2 E_{x,y}/\partial z^2$ with respect to $k_0 \partial E_{x,y}/\partial z$. Assuming a single linear polarization $E_x=E$, this allows to write:
\begin{equation}
    i\frac{\partial E}{\partial z}=-\frac{1}{2k_0}\Delta E -\frac{k_0\chi}{2}E
\end{equation}
where $\chi$ is the susceptibility. This equation is mathematically equivalent to the time-dependent Schr\"odinger equation with $z\sim t$, $k_0\sim m$, and $\chi\sim -U$ (an external potential).

The behavior of a quantum particle in a honeycomb potential is well described by the tight-binding approximation \cite{Wallace1947}. In particular, in vicinity of the corners of the Brillouin zone, the tight-binding Hamiltonian is reduced to the 2D Dirac Hamiltonian \cite{Semenoff1984}. This set of mathematical equivalences is what allows to study the Klein tunneling in this system.

\textit{Susceptibility under dual EIT effect.} In the experiment, the probe beam \bm{$E_1$} feels dual EIT window by the two coupling fields of slightly different optical frequencies: \bm{$E_2$} and \bm{$E_3$}, whilst the former creates the honeycomb lattice and the latter the barrier. The susceptibility felt by \bm{$E_1$} is expressed as follows 
\begin{equation}
    \chi=\frac{iN\left|\mu_{31}\right|^{2}}{\hbar\epsilon_{0}}\times\frac{1}{\left(\Gamma_{31}+i\Delta_{1}\right)+\frac{\left|\Omega_{2}\right|^{2}}{\Gamma_{32}+i\left(\Delta_{1}-\Delta_{2}\right)}+ \frac{\left|\Omega_{3}\right|^{2}}{\Gamma_{32}+i\left(\Delta_{1}-\Delta_{3}\right)}}
    \label{EIT}
\end{equation}
in which $N$, $\hbar$ and $\epsilon_{0}$ are the atomic density, the reduced Planck constant and the vacuum dielectric constant. $\Delta$ and $\Omega$ are the frequency detuning and the Rabi frequency for each electric field where the subscripts 1, 2 and 3 represent the probe beam \bm{$E_{1}$}, the coupling beams forming honeycomb lattice \bm{$E_{2}$} and the barrier beam \bm{$E_{3}$}. $\mu$ and $\Gamma$ are the dipole moment and decay rate, respectively, between the energy levels connected by the probe beam, where the subscripts represent the corresponding energy levels of the atoms in Fig.~\ref{fig2}(b). By choosing the correct detunings for all fields, we can obtain the proper refractive index distributions for the honeycomb lattice and the potential barrier.

The powers of the three coupling beams \bm{$E_{2}$} , \bm{$E'_{2}$} and \bm{$E''_{2}$ }(circular Gaussian beams from the same laser source and the beam diameter is $\sim$ 1.5 mm) are 15 mW, 16 mW and 30 mW, respectively, The power of the barrier beam (with the beam size of $\sim$ 1.5 mm$\times$0.3 mm) is $\sim$ 4mW. The resonant wavelength (corresponding to detuning $\Delta_{2}$= 0 and  $\Delta_{3}$= 0) of the coupling field is $\lambda_{2}$ =$\lambda_{3}$  $\approx$ 794.976 nm. The 5cm long atomic vapor cell is wrapped with µ-metal sheets to shield outside magnetic field and heated by a heat tape to 100$^{\circ}$C. The two Gaussian probe beams (with the same diameter of $\sim$1 mm) are at $\sim$0.5 mW. The resonant wavelength of the probe field corresponding to $\Delta_{1}$= 0 is $\lambda_{1}\approx 794.981$~nm.

We note that the barrier beam is co-polarized with \bm{$E_2$}. It has a slightly different two-photon detuning that induces a higher refractive index. The probe is also cross-polarized with \bm{$E_2$}. For the study of the Klein tunneling, the probe angles are tuned  within the  linear energy-momentum dispersion region ($k\ll K$, with $\bm{k}$ measured from the Dirac point $\bm{K}$). At the output of the atomic cell, the coupling beams which form the lattice and the barrier are reflected away by a PBS and the probe beams are imaged and detected by a lens and a CCD camera.

The effective height of the barrier is obtained from simulations based on the EIT equation~\eqref{EIT}, with an experimentally-set detuning $\Delta_{3}$ between -90MHz and -110MHz. The parameters of the tight-binding approximation of the honeycomb lattice $a$ (intersite distance) and $J$ (hopping rate) provide a natural reference for the energy scales involved in the experiment. The linear part of the dispersion close to the Dirac point has the scale of $J$ (the whole dispersion spans $6J$). The typical probe energy with respect to the Dirac point is thus $E\sim J$. In particular, we estimate the angular deviation of the probe beams to provide $E\approx J/2$. The simulations of the susceptibility show that the maximal barrier that can be obtained without perturbing the lattice is $V_0\approx J$. Thus, in Fig.~2, we increase the barrier up to $V_0\approx J\approx 2E$, and in Fig.~3 we scan the angular dependence at $E\approx J$ and $V_1\approx 1.1 J$, $V_2\approx 1.7 J$.

\textit{Transmission measurement.} We extract the transmission from the experimentally measured spatial distribution of intensity (Figs.~2a-f and 3a-e) by integrating the intensity in the region behind the barrier (``downstream") and dividing it by the intensity before the barrier (``upstream"). To account for the overall decrease of the intensity due to its radial redistribution as well as to other losses, we normalize the transmission to 1 by comparing it to the region without the barrier (top part of the figures). The uncertainty in the experimental figures is determined mostly by the choice of the integration regions for the calculation of the transmission.

\end{methods}

\bibliographystyle{naturemag}
\bibliography{biblio}

\begin{addendum}
\item This work was supported by National Key R\&D Program of China (2018YFA0307500, 2017YFA0303703), the Key Scientific and Technological Innovation Team of Shaanxi Province (2021TD-56), National Natural Science Foundation of China (12074303, 62022066, 12074306, 11804267). We acknowledge the support of the EU "TOPOLIGHT" project (964770), of the ANR Labex Ganex (ANR-11-LABX-0014), and of the ANR program "Investissements d'Avenir" through the IDEX-ISITE initiative 16-IDEX-0001 (CAP 20-25). The S.K. contribution was supported by the Institute for Basic Science in Korea, Young Scientist Fellowship (IBS-R024-Y3-2021)
\item[Author contributions]
Z. Zhang -- project administration, investigation, formal analysis, funding acquisition, methodlogy, visualization, investigation, writing; Y. Feng -- investigation, data curation, writing; F. Li -- conceptualization, methodology, supervision, writing, project administration; S. Koniakhin -- formal analysis, visualization; C. Li-- formal analysis; F. Liu -- formal analysis; Y. Zhang -- supervision, funding acquisition, formal analysis; M. Xiao -- supervision, formal analysis, writing; D.~Solnyshkov -- conceptualization, funding acquisition, formal analysis, methodology, visualization, writing; G.~Malpuech -- conceptualization, funding acquisition, methodology, writing, supervision.
 \item[Competing Interests] The authors declare that they have no
competing financial interests.
 \item[Correspondence] Correspondence
should be addressed to : zhyzhang@xjtu.edu.cn (Z.Z); felix831204@xjtu.edu.cn (F.L.); guillaume.malpuech@uca.fr (G.M.); dmitry.solnyshkov@uca.fr (D.S.).
\end{addendum}

\section*{Data availability statement}
The datasets generated during and/or analysed during the current study are available from the authors upon reasonable request.

\clearpage



\end{document}